\documentclass[letterpaper, 10 pt, conference]{ieeeconf}  

\IEEEoverridecommandlockouts                              
\usepackage{graphics} 
\usepackage{epsfig} 
\usepackage{mathptmx} 
\usepackage{times} 
\usepackage{amsmath} 
\usepackage{amssymb}  
\usepackage{graphicx}

\usepackage{algorithm}
\usepackage{algpseudocode}

\usepackage{subcaption}

\usepackage{booktabs} 

\usepackage{pgfplots}
\pgfplotsset{compat=1.15,
	legend style={font=\footnotesize},
}
\usepackage{tikz}
\usepackage{tikzscale}
\usepgfplotslibrary{fillbetween}

\title{\LARGE \bf
Contingency-Aware Station-Keeping Control of Halo Orbits
}

\author{Fausto Vega, Zachary Manchester, Martin Lo, and Ricardo Restrepo
\thanks{Fausto Vega and Zachary Manchester are with the Robotics Institute, Carnegie Mellon University, 5000 Forbes Ave., Pittsburgh, PA, 15213
        {\tt\small \{fvega, zmanches\}@andrew.cmu.edu}}
\thanks{Martin Lo and Ricardo Restrepo are with the California Institute of Technology Jet Propulsion Laboratory, 4800 Oak Grove Drive, Pasadena, CA 91109
         {\tt\small \{martin.w.lo, ricardo.l.restrepo\}@jpl.nasa.gov}}
         }

\begin{document}

\maketitle
\thispagestyle{empty}
\pagestyle{empty}


\begin{abstract}
We present an algorithm to perform fuel-optimal stationkeeping for spacecraft in unstable halo orbits with additional constraints to ensure safety in the event of a control failure. We formulate a convex trajectory-optimization problem to generate impulsive spacecraft maneuvers to loosely track a halo orbit using a receding-horizon controller. Our solution also provides a safe exit strategy in the event that propulsion is lost at any point in the mission. We validate our algorithm in simulations of the three-body Earth-Moon and Saturn-Enceladus systems, demonstrating both low total delta-v and a safe contingency plan throughout the mission.  

\end{abstract}


\section{INTRODUCTION}

Halo orbit missions are essential to our understanding of the solar system. Examples of such missions include the James Webb Space Telescope, an infrared telescope orbiting the Sun-Earth L2 libration point to investigate the beginnings of the universe \cite{kalirai2018scientific}, and the Genesis mission, which orbited the Sun-Earth L1 libration point gathering solar wind to explore the origins of the solar system \cite{lo2001genesis}. These orbits are periodic, highly unstable, and require several station-keeping maneuvers per orbit to track.

Howell and Pernicka introduced the target-point strategy, which minimizes a weighted cost function that penalizes deviations from the nominal orbit and control effort \cite{howell1993station}. However, the timing of the maneuvers is not optimized and the method leads to excessive fuel usage. Gomez et. al later developed a loose-control station-keeping strategy using dynamical systems theory that removed the unstable component of the invariant manifold along the orbit \cite{gomez1986station}, but this method does not optimize fuel consumption, a crucial aspect for long-term missions. Maneuver locations are typically predetermined by the mission designer and delta-v is calculated using a differential corrector. For example, the SOHO mission calculated maneuvers anywhere along the orbit using a differential corrector with the objective of making the x-component of velocity zero at the Sun-Earth line crossing in the rotating frame. \cite{dunham2001stationkeeping}. Pavlak also uses a single shooting method and compares burning at maximum y amplitudes of the orbit and four predetermined maneuver locations to show that four burns are superior in terms of fuel consumption \cite{pavlak2010mission}. However, the major drawback of shooting methods is the high dependence on the initial guess of the free variable along with the formulation of the problem to prevent ill-conditioning \cite{folta2022astrodynamics}.

Optimization-based stationkeeping was then introduced by Pavlak by formulating the station-keeping problem as a nonlinear constrained optimization problem with a quadratic objective penalizing fuel consumption \cite{pavlak2012strategy}. The quadratic objective function is not an accurate measure for fuel consumption which  leads to a continuous thrust behavior, excessive fuel consumption, and suboptimal maneuver locations. 

Contingency planning is another critical part of space missions. Due to the instability of halo orbits, the spacecraft can escape the orbit toward a planetary body if control authority is lost at any point in the mission, which raises the risk of collision with the planetary body. Since the design of halo orbits is highly flexible, it is possible to bias the initial orbit so that it will always escape into a safe orbit centered on the primary body if propulsion is completely lost.

The period of some halo orbits also poses a major problem. Usually, the navigation process at the Jet Propulsion Laboratory (JPL) requires a minimum of 24 hours to plan and perform a maneuver. However, some halo orbits have a period smaller than this window, such as the 16 hour period of an L2 halo orbit in the Saturn-Enceladus system. Therefore, autonomous solutions for maneuver planning and execution are needed. 

This paper introduces a method to optimize impulsive station-keeping maneuvers along the orbit to reduce fuel consumption and provide a safe exit strategy in the event that propulsion is completely lost during the mission. Our contributions include: 
\begin{itemize}
    \item A convex optimization approach for the station-keeping problem that minimizes fuel consumption, optimizes maneuver locations, and ensures a safe contingency plan in the event of a propulsion failure
    \item A receding-horizon control algorithm that re-solves for maneuvers twice every orbit to compensate for modeling and state-estimation errors
    \item Simulation results in the Earth-Moon and Saturn-Enceladus systems demonstrating the effectiveness of our receding-horizon controller.
\end{itemize}

The paper proceeds as follows: In Section \ref{background} we introduce the circular restricted three-body problem (CR3BP) along with concepts from dynamical systems theory used in our analysis. Section \ref{mpc} derives our trajectory optimization formulation and a receding-horizon control strategy. Closed-loop simulation results in two environments presented in Section \ref{sim_experiments}. Finally, Section \ref{conclusions} summarizes our conclusions and directions for future work.

\section{BACKGROUND}
\label{background}

This section provides a brief review of the CR3BP along with concepts from dynamical systems theory used in our analysis. We refer interested readers to \cite{koon2000dynamical} and \cite{koon2000heteroclinic} for more detailed treatments. 

\subsection{The Circular Restricted Three-Body Problem}

The CR3BP describes the motion of a small third body in the presence of two larger primary bodies. The two primary masses $m_1$ and $m_2$ are assumed to move in circular orbits about their common barycenter, and the third body is assumed to have infinitesimal mass. The mass of $m_1$ is also assumed to be larger than $m_2$, and these masses are normalized to unity. The distance between $m_1$ and $m_2$ and the rotation speed of $m_1$ and $m_2$ about the barycenter are similarly normalized to improve numerical conditioning. To reduce the time dependence of the dynamics, we use a rotating frame about the barycenter so that the two primary masses are fixed on the x-axis of the rotating frame. In this work, we performed all the analyses using two systems: the Earth-Moon system and the Saturn-Enceladus system.

We augment the CR3BP equations with a normalized thrust (acceleration) input $u = [u_x, u_y, u_z]$. The state of the spacecraft $x$ consists of its position ($q_x,q_y,q_z$) and velocity ($v_x, v_y, v_z$) in nondimensionalized units. The controlled continuous CR3BP equations of motion $\dot{\tilde{x}} = f(x, u)$ that describe the state of the spacecraft in the rotating frame are given by \eqref{cr3bp},

\begin{equation}
\begin{split}
\dot{q_x} = v_x\\
\dot{q_y} = v_y \\
\dot{q_z} = v_z \\
\dot{v_x} = \frac{\partial U}{\partial q_x} + 2v_y + u_x \\
\dot{v_y} = \frac{\partial U}{\partial q_y} - 2v_x + u_y \\
\dot{v_z} = \frac{\partial U}{\partial q_z} + u_z \\
\end{split}
\label{cr3bp}
\end{equation}
where $U$ is the augmented potential expressed in \eqref{aug_potential} and $\mu$ is the characteristic mass parameter of the CR3BP. 

\begin{equation}
\begin{split}
U = \frac{1}{2}(q_x^2 + q_y^2) + \frac{1-\mu}{r_1} + \frac{\mu}{r_2} \\
r_1 = [(q_x+\mu)^2+q_y^2+q_z^2]^{\frac{1}{2}} \\
r_2 = [(q_x-1+\mu)^2 + q_y^2 + q_z^2]^{\frac{1}{2}} \\
\end{split}
\label{aug_potential}
\end{equation}

We discretize the continuous dynamics by employing a fourth-order Runge-Kutta integrator, resulting in the discrete-time dynamics model $x_{k+1} = f_d(x_k, u_k)$, where the control input $u_k$ is discretized using a zero-order hold. This model provides the next state of the spacecraft $x_{k+1}$ given a state $x_k$ and a control $u_k$ at the time step $k$.

\subsection{Lagrange Points, Periodic Orbits, and Invariant Manifolds} 
The CR3BP has five equilibrium points, also called ``Libration'' or ``Lagrange'' points and commonly numbered L1-L5. The unstable Libration points, L1-L3, are surrounded by low-energy unstable orbits, known as ``halo orbits,'' where very small perturbations lead to large deviations from the orbit. These orbits are useful for mission design, as they maximize observation efficiency and minimize environmental impacts \cite{bauer2002libration}. However, halo orbits require satellites to execute frequent station-keeping maneuvers due to instability. We generated an unstable halo orbit about L2 using a differential correction with a third-order Richardson expansion as an initial guess \cite{koon2000dynamical}; however, obtaining halo orbit initial conditions is also possible via the JPL Solar System Dynamics site \cite{jpl_periodic_orbits}. This generated orbit is then used to form the reference trajectory that the satellite tracks.

\begin{figure}[H]
\centering
\begin{subfigure}{\columnwidth} 
\centering
\resizebox{0.8\linewidth}{!}{\input{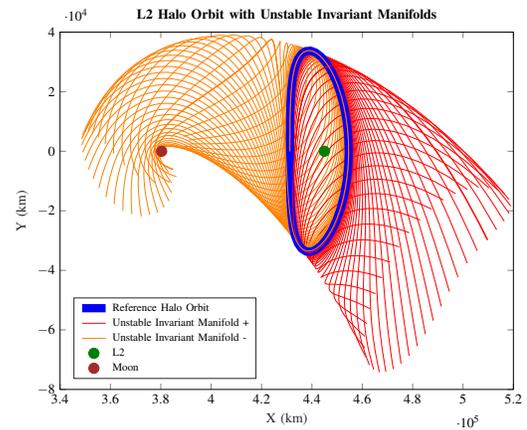}}
\caption{Halo orbit around L2 in the Earth-Moon system with a period of 14.81 days. Adding the perturbation results in the unstable invariant manifold on the right, and subtracting the perturbation leads to the unstable manifold on the left.}
\label{earth-moon-L2}
\end{subfigure}
\par\medskip 
\begin{subfigure}{\columnwidth} 
\centering
\resizebox{0.8\linewidth}{!}{\input{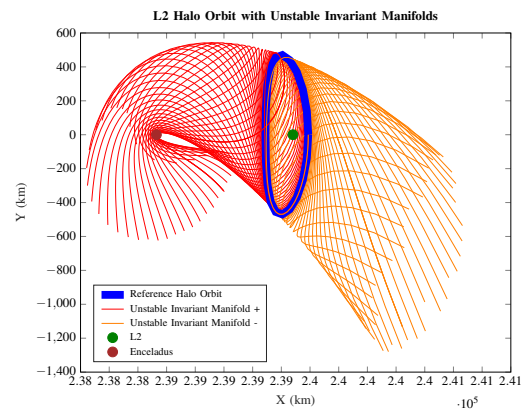}} 
\caption{Halo orbit around L2 in the Saturn-Enceladus system with a period of 16.21 hours. Adding the perturbation results in the unstable invariant manifold on the left, and subtracting the perturbation leads to the right unstable manifold}
\label{saturn-enceladus-L2}
\end{subfigure}
\caption{Reference halo orbits in the Earth-Moon and Saturn-Enceladus system along with their unstable invariant manifolds that we use for a safe exit.}
\label{reference_orbits}
\end{figure}

Along these orbits, there exist sets of low-energy trajectories that escape and approach the periodic orbit, known as invariant manifolds. These manifolds are tube-like structures that are used for various mission-design applications, such as the design of gravity-assist flybys, low-energy capture, and escape around bodies in the solar system. To obtain unstable manifolds, we first calculate the state-transition matrix $\Phi$ at each timestep $k$ by integrating the matrix differential equation \eqref{stm}. 
 \begin{equation}
 \begin{split}
     \dot{\Phi}(t) =  \frac{\partial f}{\partial x} \Phi(t) \\
     \Phi(0) = I
     \label{stm}
\end{split}
 \end{equation}
 The state-transition matrix integrated over one orbit period $T$ is known as the monodromy matrix, 
 \begin{equation}
M = \Phi(T) .
\end{equation}
The unstable eigenvector of $M$, which we label $v$, is locally tangent to the unstable manifold. We use this unstable direction to generate initial conditions for trajectories $w$ on the unstable manifold and then integrate the continuous dynamics $f(x, u)$ forward to time $\tau$ using a higher-order integrator. The computation of these trajectories is summarized in \eqref{unstable_manifold}, where $\varepsilon$ is a parameter that scales the magnitude of the perturbation in the unstable direction. 
 \begin{gather}
w (t) = \int f (x (t) \pm \varepsilon \Phi (t) v, 0) d\tau \quad
t \in [0, T]
\label{unstable_manifold}
\end{gather}

Figure \ref{reference_orbits} shows example halo orbits to be tracked by our method, along with their unstable manifolds.

\section{Optimization-Based Stationkeeping}
\label{mpc}

This section derives our trajectory optimization formulation and receding-horizon control strategy to solve for fuel optimal stationkeeping maneuvers. We also describe the linearized dynamics used in our convex optimization problem. 
\subsection{Linearized Dynamics}
We discretize the reference halo orbit into $N$ knot points.
Using a first-order Taylor expansion, we linearize the nonlinear discrete-time dynamics at each knot point $\Bar{x}_k$, leading to the linear state error dynamics:

 \begin{equation}
\begin{split}
\Delta x_{k+1} \approx A_k\Delta x_k + B_k  u_k  \\
A_k =  \frac{\partial f_d}{\partial x_k}|_{\Bar{x}_k, u_k} \ \ \ \ B_k = \frac{\partial f_d}{\partial u_k}|_{\Bar{x_k}, u_k} \\
\Delta x_k = x_k - \Bar{x}_k \\
\end{split}
\label{firstorder-taylor}
\end{equation}

where $A_k$ and $B_k$ are the discrete dynamics Jacobians evaluated at the reference trajectory at timestep $k$.

\subsection{Trajectory Optimization}
We pose the station-keeping problem as a trajectory optimization problem in \eqref{convex_prob}, where $l_k(x_k, u_k)$ is the stage cost, $l_N(x_k, u_k)$ is the terminal cost, $g(x_k, u_k)$ are the equality constraints and $h(x_k, u_k)$ are the inequality constraints. We solve the problem over a 2 orbit horizon $2N$, where $N$ is the number of discrete knot points along the orbit. Convex cost and constraint functions are chosen to guarantee that the solution is globally optimal and can be solved efficiently \cite{boyd2004convex}.


\begin{equation}
\begin{aligned}
& \underset{x_{1:2N}, u_{1:2N-1}}{\text{minimize}}
& & J = \sum_{k}^{2N-1} l_k(x_k,u_k) + l_N(x_k,u_k)\\
& \text{subject to}
& & g(x_k, u_k) = 0 \; \\
&&& h(x_k, u_k) \leq 0 \; \\
\end{aligned}
\label{convex_prob}
\end{equation}

We focus on minimizing fuel consumption along the orbit while satisfying mission objectives and constraints. The mission objectives include completing a certain number of revolutions around a halo orbit to obtain a large number of observations while minimizing fuel consumption. We chose to minimize the L1 norm of the thrust vector because $u_k$ is delta v in discrete time, and the sum of delta v across a horizon approximates the total fuel consumption. The L1 norm also encourages impulsive solutions which is desired when minimizing fuel \cite{ross2004find}:
\begin{equation}
    l(x_k, u_k) = ||u_k||_1
    \label{l1_cost}
\end{equation}

Instead of a tracking cost, we employ constraints on maximum state error. Two variants of the state constraint were tested: The first constraint is a Euclidean ball constraint $D$, which restricts the deviation from the reference trajectory $\Delta x$ to a user-defined radius as shown in \eqref{tube-constraint}. The state radius $r$ consists of the position limit $r_q$ and the velocity limit $r_v$, which are determined depending on how closely the user wants the satellite to follow the reference trajectory.
This constraint will allow the spacecraft to drift from the reference trajectory and only provide a control impulse when necessary. 

\begin{equation}
\begin{split}
    D(\Delta x_k, r) = \{\Delta x_k \ | \ ||\Delta x_k||_2 \leq r\} \\
    r = \begin{bmatrix}
    r_q \\
    r_v
        \end{bmatrix}
    \label{tube-constraint}
\end{split}
\end{equation}

\begin{figure}
\centering
\begin{subfigure}{\columnwidth} 
\centering
\includegraphics[width=0.7\textwidth]{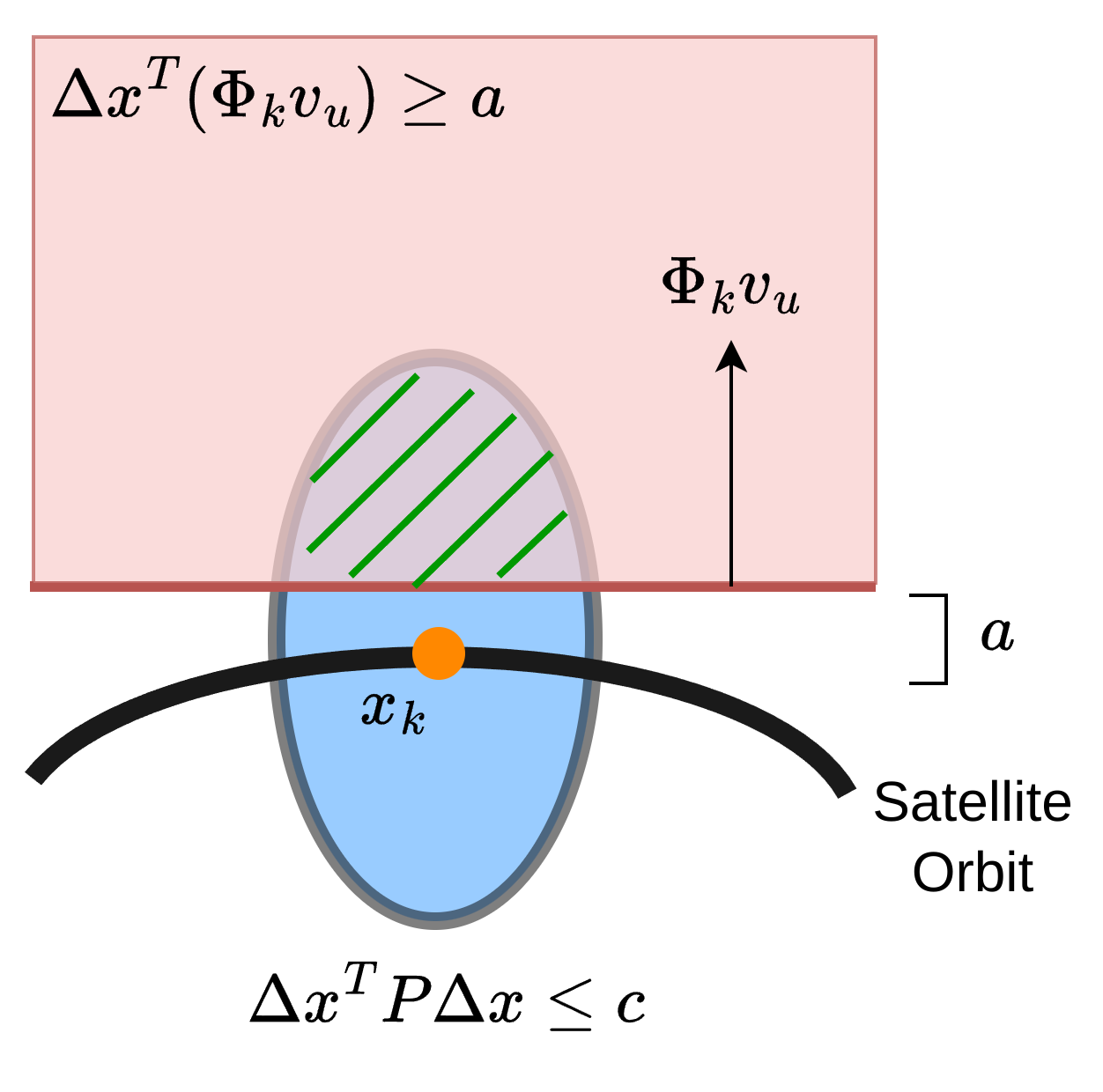} 
\caption{Ellipsoid cost-to-go constraint with the invariant manifold half-space constraint}
\label{ellipsoid-constraint-2d}
\end{subfigure}
\par\medskip 
\begin{subfigure}{\columnwidth} 
\centering
\includegraphics[width=0.7\textwidth]{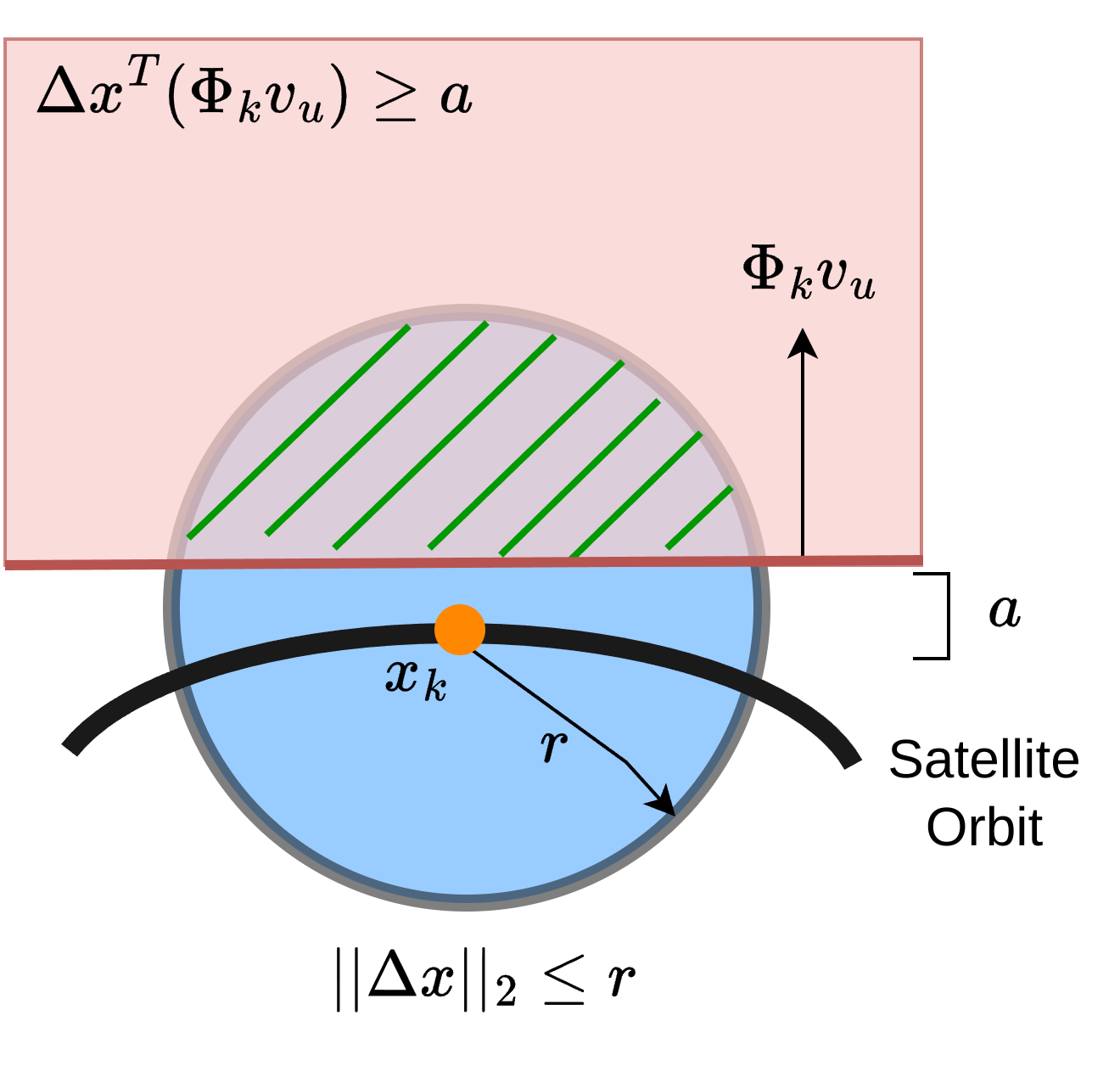} 
\caption{Euclidean ball constraint with the invariant manifold half-space constraint}
\label{tube-constraint-2d}
\end{subfigure}
\caption{2D projection of the state constraints along the orbit $x_k$. The feasible set is the area intersected with green lines.}
\label{state-constraints}
\end{figure}

   
The second constraint is an ellipsoid determined by taking a level set of the quadratic cost-to-go function computed by solving a Linear-Quadratic Regulator (LQR) tracking problem. This state constraint is better informed by the dynamics of the problem, therefore we expect better performance compared to the Euclidean ball constraint. Since the orbit is periodic, the cost-to-go is also periodic for this system \cite{bittanti1991periodic}. In \eqref{lqr-cost}, $Q_k$ and $R_k$ are cost weighting matrices for the states and controls, while $Q_N$ is the terminal cost weighting matrix.  The periodic cost-to-go term was calculated by solving the Ricatti recursion until convergence to a periodic solution \cite{underactuated}.


\begin{equation}
J  = \sum_{k=1}^{2N-1} \frac{1}{2} x_k^T Q_k x_k + \frac{1}{2} u_k^T R_k u_k + \frac{1}{2} x_N^T Q_N x_N
\label{lqr-cost}
\end{equation}



\begin{equation}
    E(\Delta x_k, P_k) = \{\Delta x_k \ | \ \Delta x_k^T P_k \Delta x_k \leq c \}
    \label{ctg-constraint}
\end{equation}

Next, an initial-state equality constraint is imposed to set the initial delta state to reflect the current observed state of the system $x_0$. 
 \begin{equation}
 \Delta x_1 = \Delta x_0
 \label{initial_state}
\end{equation}

\subsection{Contingency Constraint}
Lastly, we impose a half-space constraint $M$ on the state deviation from the reference trajectory $\Delta x_k$ to ensure that it contains some component in the unstable manifold that escapes away from the planetary body. As long as there is some component in the unstable direction, it will leave through the unstable manifold described in \cite{vegamassively}. This constraint allows the spacecraft to bias its trajectory toward the unstable manifold to safely exit the halo orbit if the spacecraft completely loses thrust during the mission. This capability allows mission designers to re-plan the mission in safe mode if thrust is regained at a later time. This exit strategy also prevents collisions with planetary bodies. For example, in the Earth-Moon system, we can ensure a safe exit through the right of the L2 point to avoid the Moon, which is located on the left of the L2 point. In \eqref{manifold_constraint}, $\Phi_k$ is the state transition matrix at time step k, and multiplying the unstable direction $v_u$ propagates this unstable direction to any time step k. The direction of the exit trajectory (either left or right) is determined by the sign of $v_u$ as illustrated in Fig. \ref{unstable_manifold}; however, this does not hold for a small subset of directions due to the chaotic nature of the dynamics. A visual representation of the ellipsoidal, Euclidean ball, and half-space constraint is shown in Fig. \ref{state-constraints}. 
\begin{equation}
M(\Delta x_k, \Phi_k) = \{\Delta x_k \ | \ \Delta x_k^T (\Phi_k v_u) \geq a \}
\label{manifold_constraint}
\end{equation}

Our full problem formulation is,
\begin{equation}
\begin{aligned}
& \underset{u_{1:2N-1}}{\text{minimize}}
& & J = \sum_{k=1}^{2N-1} ||u_k||_1\\
& \text{subject to}
& & \eqref{firstorder-taylor}, \ \eqref{initial_state} \; \\
&&& \eqref{tube-constraint} \ \text{or} \ \eqref{ctg-constraint}, \ \eqref{manifold_constraint} \; \\
\end{aligned}
\label{convex_prob_final}
\end{equation}
\eqref{convex_prob_final}, where the equality constraints are the linearized discrete-time error dynamics and the inequality constraints consist of a Euclidean ball or ellipsoidal state constraint and an unstable direction half-space constraint.




\section{SIMULATION EXPERIMENTS}
\label{sim_experiments}

\begin{table}[h]
        \centering
        \caption{Simulation Constants}
         \begin{tabular}{lll}
            \addlinespace[1ex]
            Constant & Earth-Moon & Saturn-Enceladus \\
            \addlinespace[1ex]
            \midrule[1pt]
            \addlinespace[1ex]
            $\mu$ & 1.215 $\times 10^{-2}$ & 1.901 $\times 10^{-7}$ \\
            LU [km]  & 3.850 $\times 10^5$ & 2.38529 $\times 10^5$ \\
            TU [days] & 4.349 & 0.2189\\
            \addlinespace[1ex]
            \bottomrule[1pt] 
            \end{tabular}
        \label{norm-constants}
\end{table}


We simulated 100 revolutions around the L2 halo orbit shown in Fig. \ref{earth-moon-L2}. The CR3BP dynamics are nondimensionalized by the constants in Table \ref{norm-constants}. To solve the optimization problem in \eqref{convex_prob_final}, we use Convex.jl \cite{udell2014convex}, a convex optimization modeling framework in Julia \cite{bezanson2012julia}, and the Mosek solver \cite{aps2024mosek}. To simulate the nonlinear CR3BP dynamics, we use a higher-order integrator, specifically the Tsitouras-Papakostas 8/7 Runge-Kutta method in DifferentialEquations.jl \cite{rackauckas2017differentialequations}. 
To mitigate linearization errors, we implemented a receding-horizon controller that solves the optimization problem in \eqref{convex_prob_final} for a two-revolution horizon. We then simulated half of an orbit of optimal controls on the discrete nonlinear dynamics to obtain the new initial state for the next solution. This iterative process is then repeated for the specified mission time. The output of the solver is thrust commands; therefore, we approximate fuel consumption as delta v which is approximated as $u_k \Delta t$. We validate our algorithm in two different settings to assess its efficiency in computing fuel-efficient maneuvers and devising a secure contingency plan in case of losing control. 


\begin{figure}
\centering
\begin{subfigure}{\columnwidth} 
\centering
\resizebox{\linewidth}{!}{\input{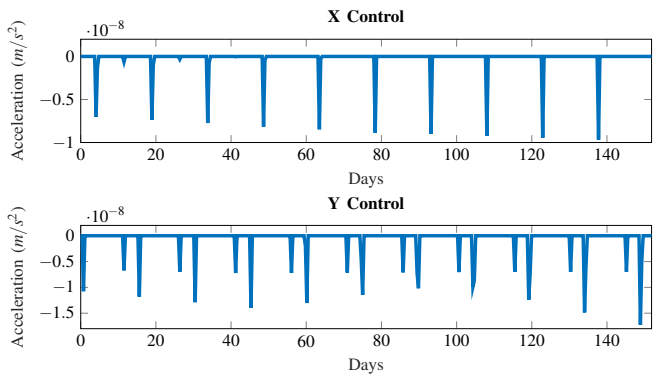}}
\caption{Euclidean ball state constraint control strategy}
\label{tube-saturn-enceladus}
\end{subfigure}
\par\medskip 
\begin{subfigure}{\columnwidth} 
\centering
\resizebox{\linewidth}{!}{\input{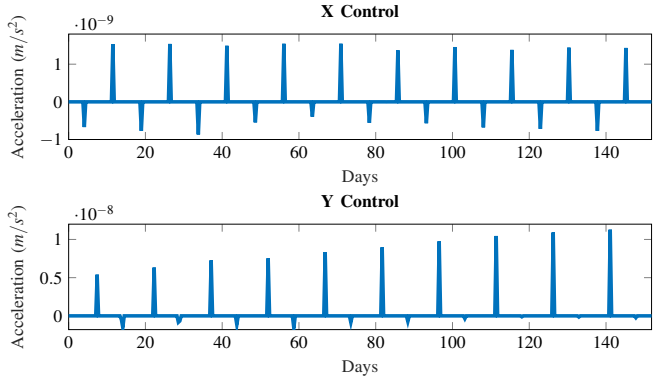}} 
\caption{Cost-to-go (ellipsoidal) state constraint control strategy}
\label{ctg-saturn-enceladus}
\end{subfigure}
\caption{Comparison of two control strategies for revolutions 10-20 of an L2 halo orbit in the Earth-Moon system. The z control is zero for both strategies.}
\label{control-earth-moon}
\end{figure}

\subsection{Earth-Moon Simulation}

For this experiment, the initial delta state condition also know as "injection error" was set to 385 $m$ in the x position and 1.856 $\frac{m}{s}$ in the y-component of the velocity. The goal is to loosely follow the orbit from Fig. \ref{earth-moon-L2} and ensure a safe departure trajectory throughout the entire mission in the event of an emergency. The halo orbit was discretized into 41 knot points, which resulted in a timestep $\Delta t = 8.911$ hours. First, we compare the fuel consumption between the Euclidean ball and the ellipsoidal constraints. The state weighting matrices $Q_k$ and $Q_N$  for the cost-to-go calculation were set to $1 \times 10 ^ {-3} I(6)$ where $I(6)$ is a 6 $\times$ 6 identity matrix, the control weight matrix $R_k$ was set to $1 \times 10 ^ {3} I(3)$, and $c=1 \times 10^4$ in \eqref{ctg-constraint}. This cost-to-go metric penalizes control more than state deviation, which matches our fuel consumption objective. For the Euclidean ball constraint, the radius for position $r_q$ and velocity $r_v$ was set to 1000 which will allow the spacecraft to deviate from the reference and only burn when necessary. The constant $a = 1 \times 10^{-2}$ in \eqref{manifold_constraint}, and this value was tuned to ensure that the manifold constraint was satisfied throughout the entire trajectory. The results of the control inputs for revolutions 10-20 are shown in Fig. \ref{control-earth-moon}, where the top plot represents the control with the Euclidean ball constraint and the bottom plot shows the control with the ellipsoid state constraint. For the 100 revolutions, the Euclidean ball constraint consumed 2.89 $\frac{m}{s}$ of fuel while the ellipsoidal constraint used 2.713 $\frac{m}{s}$. A majority of the fuel is consumed to correct the initial injection error as only 0.357 $\frac{m}{s}$ was consumed for revolutions 2-100 using the Euclidean ball constraint, while the ellipsoid constraint only used 0.0908 $\frac{m}{s}$ for revolutions 2-100. 

The total fuel consumption per year is shown in Table \ref{fuel-consumption}. As expected, the cost-to-go ellipsoid constraint uses less fuel because the state space is constrained using the LQR cost metric, which heavily penalizes the control effort. The manifold constraint is satisfied throughout the entire 100 revolution trajectory, and to verify that the satellite leaves through the right unstable manifold if the thrusters were to malfunction, we simulate CR3BP dynamics without control forward in time at each state of the solution. As expected, the satellite escapes through the right manifold throughout the mission, as the optimizer provided stationkeeping maneuvers that autonomously biased the trajectory in the unstable manifold direction. However, due to the injection error, there is some transient phase in which the manifold constraint does not guarantee a safe exit. This is because the spacecraft initially executes thrust maneuvers to reduce the effects of the injection error, and then it reaches a steady-state phase where it guarantees a safe exit trajectory. The safe exit trajectory for revolutions 3-100 Euclidean ball constraint is shown in Fig. \ref{manifold-exit-earthmoon}.

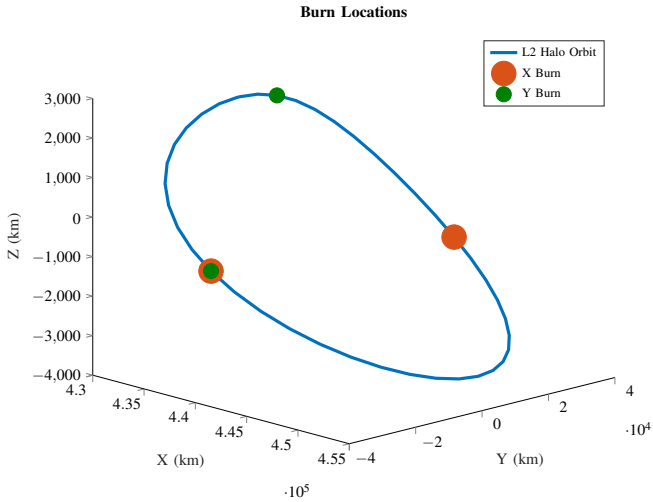
\begin{figure}[h]
    \centering
    \resizebox{\linewidth}{!}{
%
%
\definecolor{mycolor1}{rgb}{0.00000,0.44700,0.74100}%
\definecolor{mycolor2}{rgb}{0.85000,0.32500,0.09800}%
\definecolor{mycolor3}{rgb}{0.92900,0.69400,0.12500}%
\begin{tikzpicture}

\begin{axis}[%
width=4.568in,
height=3.603in,
at={(0.766in,0.486in)},
scale only axis,
xmin=430000,
xmax=455000,
tick align=outside,
xlabel style={font=\color{white!15!black}},
xlabel={X (km)},
ymin=-40000,
ymax=40000,
ylabel style={font=\color{white!15!black}},
ylabel={Y (km)},
zmin=-4000,
zmax=3000,
zlabel style={font=\color{white!15!black}},
zlabel={Z (km)},
view={46}{19},
axis background/.style={fill=white},
title style={font=\bfseries},
title={Burn Locations},
axis x line*=bottom,
axis y line*=left,
axis z line*=left,
legend style={legend cell align=left, align=left, draw=white!15!black}
]
\addplot3 [color=mycolor1, line width=2.0pt]
 table[row sep=crcr] {%
431249.946141646	0	2286.76971698967\\
431267.033059025	5775.21427993587	2237.80569344136\\
431338.250449522	11343.5921017774	2094.63198106384\\
431515.860687375	16521.6569602332	1867.45545115446\\
431866.202834045	21163.3375409244	1570.66285393354\\
432450.344561395	25162.1803368925	1220.27439314907\\
433310.590521555	28445.7013908956	832.013140162874\\
434464.506635278	30967.3811478107	420.233325666567\\
435904.597835624	32699.9235665503	-2.46904429596334\\
437601.144187165	33630.9097151082	-425.242762640362\\
439506.285654924	33760.6574103307	-838.791214671086\\
441558.385846515	33101.5547168056	-1235.1338658286\\
443686.178393286	31678.2658126305	-1607.39135135693\\
445812.69496354	29528.2389615181	-1949.61064452896\\
447858.948047565	26702.1531065535	-2256.63747497383\\
449747.451253236	23264.156080221	-2524.03650661161\\
451405.554273138	19291.5474679803	-2748.0431656137\\
452768.653453927	14874.0351037452	-2925.55177663502\\
453783.01516271	10112.2222358645	-3054.11576337949\\
454408.343753312	5115.62280147704	-3131.96876008347\\
454619.602600862	4.38578368955789e-05	-3158.03893004773\\
454408.343337223	-5115.62239346339	-3131.96873991848\\
453783.016015575	-10112.2224823329	-3054.11580067173\\
452768.652164753	-14874.0338416253	-2925.55170308197\\
451405.555712527	-19291.5487034221	-2748.04324726804\\
449747.449903355	-23264.1532789562	-2524.03637099744\\
447858.949275939	-26702.1555739773	-2256.63760658047\\
445812.695049319	-29528.2360429389	-1949.61049580817\\
443686.178283345	-31678.2706506109	-1607.39163647731\\
441558.386874359	-33101.5534846914	-1235.13377845568\\
439506.286441207	-33760.6578773395	-838.791222033601\\
437601.140635793	-33630.9139356952	-425.243182326914\\
435904.600668232	-32699.9228777203	-2.46888339633341\\
434464.511562599	-30967.3795214029	420.233625348347\\
433310.59390317	-28445.7009068327	832.01328927701\\
432450.344199524	-25162.1834259942	1220.27428617782\\
431866.205154381	-21163.3376849873	1570.66285361715\\
431515.862232943	-16521.65949159	1867.45544690779\\
431338.252114162	-11343.5939270574	2094.6318909861\\
431267.034259993	-5775.2166343082	2237.80547333561\\
431249.949366846	-0.00385146398065033	2286.76979583685\\
};
 \addlegendentry{L2 Halo Orbit}

\addplot3[only marks, mark=*, mark options={}, mark size=7.7386pt, color=mycolor2, fill=mycolor2] table[row sep=crcr]{%
x	y	z\\
441558.385846515	33101.5547168056	-1235.1338658286\\
439506.286441207	-33760.6578773395	-838.791222033601\\
};
\addlegendentry{X Burn}

\addplot3[only marks, mark=*, mark options={}, mark size=4.7678pt, color=black!50!green, fill=black!50!green] table[row sep=crcr]{%
x	y	z\\
431338.250449522	11343.5921017774	2094.63198106384\\
439506.286441207	-33760.6578773395	-838.791222033601\\
};
\addlegendentry{Y Burn}

\end{axis}
\end{tikzpicture}%
}
    \caption{Burn locations for the L2 halo orbit in the Earth Moon system using the Euclidean ball state constraint}
    \label{burn_locations-earth-moon}
\end{figure}

We also analyzed the burn locations along the orbit for one of the scenarios.  Fig. \ref{burn_locations-earth-moon} shows the location of the burns for the control trajectory from the Euclidean ball constraint solution. Interesting patterns emerge, such as the symmetry of the x burns around the y-z plane and the simultaneous x and y burns in the same instance.


 \begin{table}[h]
        \centering
        \caption{Fuel Consumption per Year}
         \begin{tabular}{lll}
            \toprule[1pt]
            \addlinespace[1ex]
            System & State Constraint & Fuel Consumption [$\frac{m}{s}/yr$] \\
            \addlinespace[1ex]
            \midrule[1pt]
            \addlinespace[1ex]
            Earth-Moon & Euclidean Ball & 0.712\\
            Earth-Moon  & Ellipsoid & 0.668\\
            Saturn-Enceladus & Euclidean Ball & 30.16\\
            Saturn-Enceladus  & Ellipsoid & 28.755\\
            \addlinespace[1ex]
            \bottomrule[1pt] 
            \end{tabular}
        \label{fuel-consumption}
\end{table}

\subsection{Saturn-Enceladus Simulation}
\begin{figure}
\centering
\begin{subfigure}{\columnwidth} 
\centering
\resizebox{\linewidth}{!}{\input{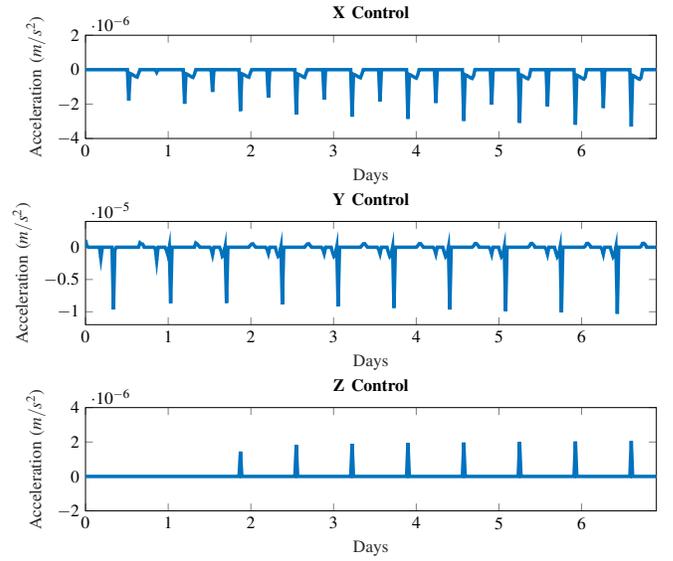}}
\caption{Euclidean ball state constraint control strategy}
\label{tube-saturn-enceladus}
\end{subfigure}
\par\medskip 
\begin{subfigure}{\columnwidth} 
\centering
\resizebox{\linewidth}{!}{\input{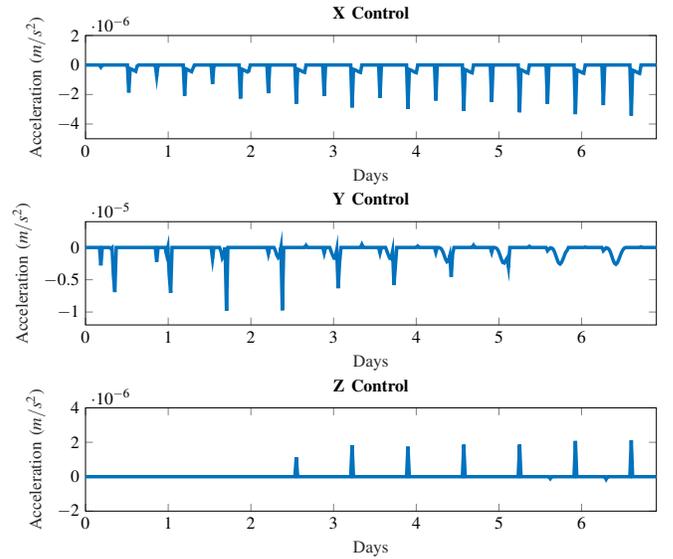}} 
\caption{Cost-to-go state (ellipsoidal) constraint control strategy}
\label{ctg-saturn-enceladus}
\end{subfigure}
\caption{Comparison of two control strategies for revolutions 10-20 of an L2 halo orbit in the Saturn-Enceladus system.}
\label{control-saturn-enceladus}
\end{figure}

For the Saturn-Enceladus system, the initial injection error was set to 238.5 $m$ in the x position and 0.486 $\frac{m}{s}$ in the y-component of the velocity. Similarly, the goal is to track the halo orbit from Fig. \ref{saturn-enceladus-L2} with minimal fuel and ensure a safe exit trajectory throughout most of the mission. The L2 halo orbit was discretized to 41 knot points, resulting in a timestep $\Delta t = 24.308$ minutes. The state weighting matrices $Q_k$ and $Q_N$  for the cost-to-go calculation were set to $1 \times 10 ^ {-6} I(6)$, the control weight matrix $R_k$ was set to $1 \times 10 ^ {-3} I(3)$, and $c=1$ in \eqref{ctg-constraint}. These values are significantly smaller than those of the Earth-Moon system; however, they were chosen to maintain good numerical stability, and they still penalize the control usage more than the state deviation. For the Euclidean ball constraint, $r_q$ and $r_v$ were set to 100 and the constant $a$ was set to $5 \times 10^{-1}$ in \eqref{manifold_constraint}. The results of the control inputs for revolutions 10-20 are depicted in Fig. \ref{control-saturn-enceladus}, and the fuel consumption for the Euclidean ball constraint was 5.586 $\frac{m}{s}$ while the ellipsoidal constraint used 5.235 $\frac{m}{s}$ for 100 revolutions. 

\begin{figure}
\centering
\begin{subfigure}{\columnwidth} 
\centering
\includegraphics[width=\textwidth]{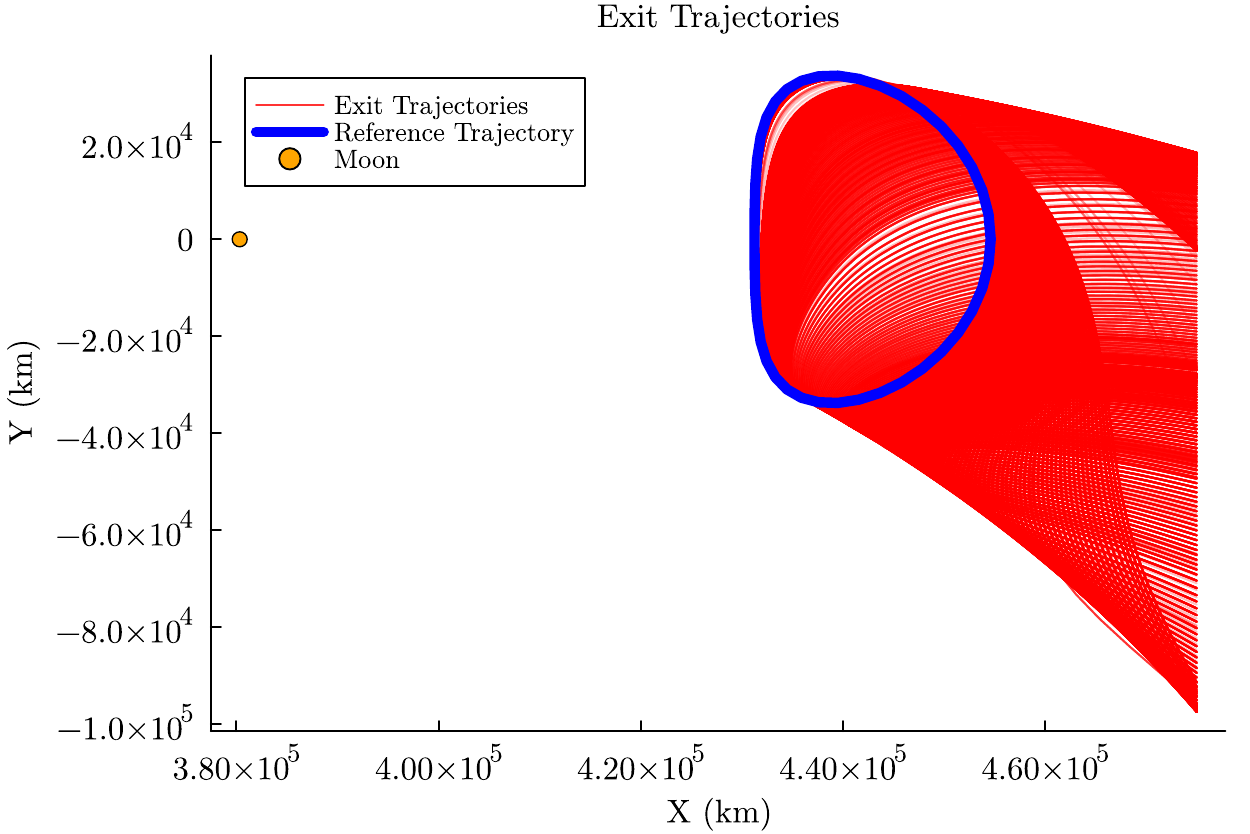} 
\caption{Manifold exit trajectories for the Earth-Moon solution with the Euclidean ball constraint for revolutions 4-100 (steady state phase). We obtain a 99.92\% success rate in generating safe exit trajectory throughout the entire 100 revolutions.}
\label{manifold-exit-earthmoon}
\end{subfigure}
\par\medskip 
\begin{subfigure}{\columnwidth} 
\centering
\includegraphics[width=\textwidth]{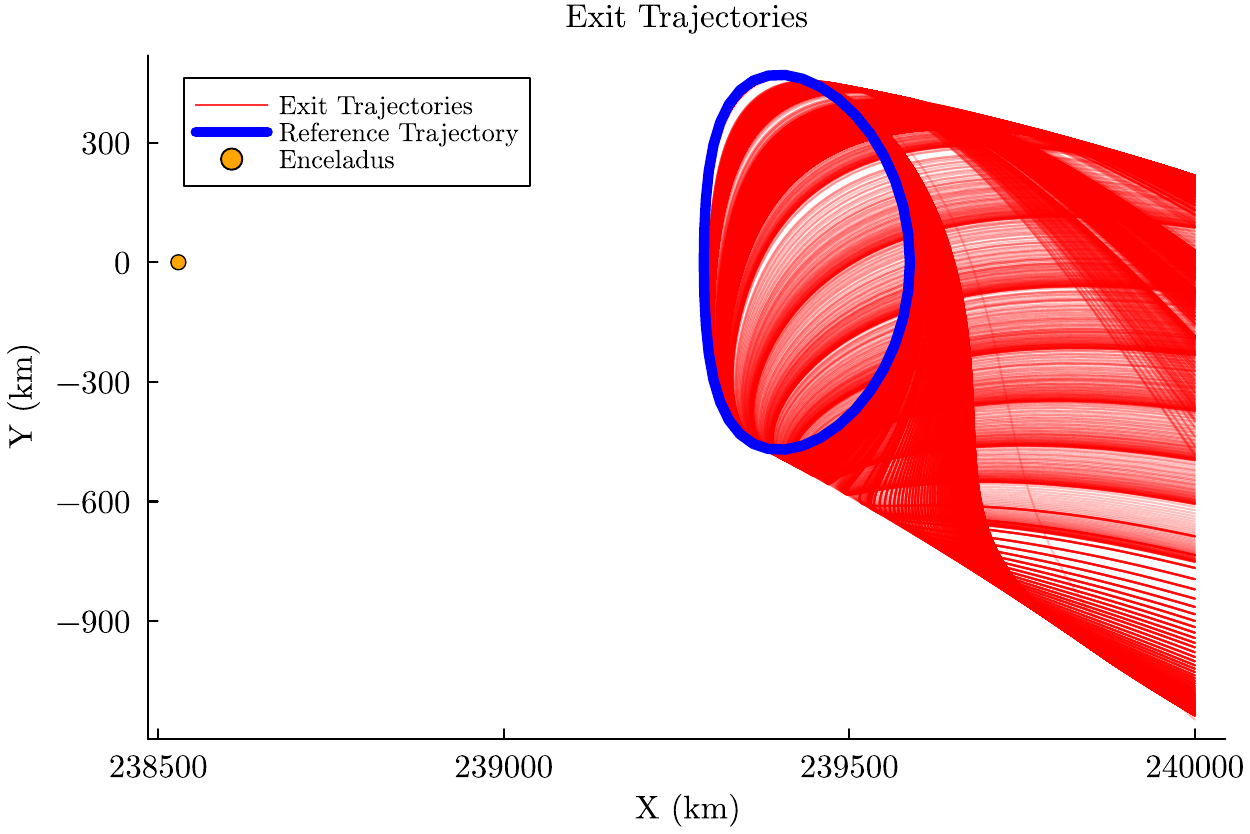} 
\caption{Manifold exit trajectories for the Saturn-Enceladus solution with the Euclidean ball constraint for revolutions 12-100 (steady state phase). We obtain a 97.53\% success rate in generating a safe exit trajectory throughout the entire 100 revolutions.}. 
\label{manifold-exit-saturn-enceledus}
\end{subfigure}

\label{manifold-exit}
\end{figure}

The fuel consumption per year is listed in Table \ref{fuel-consumption}. Again, the ellipsoidal constraint is superior by a smaller margin, and both constraints provide impulsive burns, which is desired. For the manifold escape trajectory in the Saturn Enceladus system, we simulated the uncontrolled dynamics along all states from the problem using the Euclidean ball constraint solution and the satellite successfully exited through the right unstable manifold for all states in revolutions 12-100 as shown in Fig. \ref{manifold-exit-saturn-enceledus}. Since the dynamics in the Saturn-Enceladus case are significantly faster than the Earth-Moon, the transient phase lasts for a longer period, however, we still obtain a safe exit trajectory for revolutions 12-100.


\section{CONCLUSIONS}
\label{conclusions}
In this paper, we studied the use of a receding-horizon controller for long-term stationkeeping around an unstable halo orbit that offers a secure contingency plan in the event of thruster malfunctions. The exit strategy is devised by introducing a half-space constraint in the optimization problem to bias the satellite trajectory in a desired direction on the unstable manifold. The controller produces impulsive thrusts at the optimal locations around the orbit and minimizes fuel consumption throughout the desired number of revolutions.
Future work includes simulating the dynamics using an accurate N-body dynamics model with ephemeris data and introducing model uncertainty such as state estimation error and maneuver execution error to provide guarantees using robust model-predictive control. 

      

\section{ACKNOWLEDGMENTS}

This material is based upon work supported by the National Science Foundation under Grand No. DGE2140739. 
 The research was carried out at the Jet Propulsion Laboratory, California Institute of Technology,
 under a contract with the National Aeronautics and Space Administration (80NM0018D0004).
The research was supported in part by the Jet Propulsion Strategic University Research
Partnership Program.


\bibliographystyle{IEEEtran}
\bibliography{bibliography/references}

\end{document}